\documentstyle[twocolumn,prl,aps]{revtex}
\begin{document}
\title{ Comment on ``Renormalized Coupling Constant for
the Three-Dimensional Ising Model''}
\maketitle
{In a recent Letter\cite{BAK}, Baker and Kawashima (BK) reported their
Monte Carlo result on the renormalized
coupling constant of the three-dimensional (3D) Ising model,
and claimed that their data verify the hyperscaling relation~(HR) 
for this model.
We found, however, that the data and analysis in the Letter
do not verify what the authors claim, and that the Letter contains
no significant new findings.

The thermodynamic value of the four-point renormalized coupling 
constant defined at zero
momentum has a well defined scaling behavior in terms of the
inverse temperature $K$, given by
$g(K) \sim (K_{c}-K)^{D\nu-2\Delta+\gamma}$. \\
For the 3D $\lambda \phi^{4}$ model 
it has been rigorously  proved\cite{GLM} that HR, 
\begin{equation}
D\nu-2\Delta+\gamma=0,
\end{equation}
holds for small values of $\lambda$. 
Note that HR simply means that
$g(K)$ remains a non-zero constant in the scaling regime 
in the absence of a multiplicative correction to scaling.

For the 3D Ising model which corresponds to the 
3D $\lim_{\lambda \to \infty} \lambda \phi^{4}$ model,
Kim and Patrascioiu (KP)\cite{KIM} 
have already found using Monte Carlo method that $g(K)$ 
remains a constant in the scaling regime i.e., 
$g(K) \simeq 25(1)$ over a broad range of the 
correlation length $ 4.45(2) \le \xi \le 14.53(5)$. 
The value of $g(K)$ is in agreement with the result
calculated by a different method\cite{MAX}.
Notice that the largest value of the $\xi$ BK considered is just
6.58, being much smaller than that of KP.
More crucially, most of the data in the Letter are not actually
in the scaling regime in the sense that 
their values of $\xi(K)$ are not sufficiently large 
as compared to the lattice spacing;
this is why they have observed 
the strong temperature dependence of $g(K)$ (Table(1)).
Namely, their $g(K)$ decreases monotonically as $K \to K_{c}$ 
(from $g=45.4$ to 25.5), 
which would rather indicate $D\nu-2\Delta+\gamma >0$.

In the Letter the authors have confused $g(K_{c})$ for
the thermodynamic value of $g(K)$ in the scaling regime 
(which was denoted by 
$g^{*}\equiv \lim_{K \to K_{c}} g(K)$).
In order to obtain a thermodynamic value for a given 
$K$ through Monte Carlo simulation, 
the linear size of the lattice (L) should be much larger 
than the corresponding correlation length (thermodynamic condition),
e.g., $L/\xi(K) \ge 5$ for the 3D Ising model\cite{KIM}\cite{MAX}.
At criticality one can never satisfy the thermodynamic condition,
so that $g(K_{c})$ can never equal $g^{*}$.
The comparison between the two is simply meaningless,
and ``the significant amount of difference between these
two results''\cite{BAK} is not surprising. 

BK contend that $g(K_{c})$ is the lower bound of $g^{*}$, and
that $g(K_{c})>0$ is a verification of HR. 
This contention is a direct result of their confusion of
$g(K_{c})$ for $g^{*}$, and thus it cannot be justified.
An example against this contention can be taken from the
2D Ising model where the the hyperscaling, Eq.(1), 
has been {\it rigorously} proven\cite{AIZ}; 
at the same time, there has been a consensus on both the 
thermodynamic value $g^{*}$ and $g(K_{c})$ for this model.
All the calculations based on different methods, 
including Monte Carlo, series expansion, 
and field theoretic renormalization group,
have yielded $g^{*}=14.2(5)$, whereas $g(K_{c})=2.24(1)$\cite{KIM}. 
These two values are definitely different, showing that
the two values are actually discontinuous.
Even when HR is violated, meaning that $g^{*}=0$, 
$g(K_{c})$ will vary so slowly with L 
that its value will always remain positive on any finite lattice. 
$g(K_{c})$ is strictly independent of L for its sufficiently large 
value only if HR holds\cite{COM}; this independence of L was
also confirmed by KP for the 2D and 3D Ising models\cite{KIM}. 

BK took $L/\xi(K)\simeq 10$ for their measurement 
of the thermodynamic $g(K)$.
This is unnecessary, as can be easily checked by comparing the 
overlapping data by BK and KP near $K=0.218$ (see Table(1) both
in the Letter and [3]): clearly, 
$g$ measured under the condition $L/\xi(K) \simeq 6$ is 
consistent with the result of BK, showing that
the thermodynamic condition for the 3D Ising model 
is indeed satisfied for $L/\xi(K) \simeq 6$\cite{KIM}\cite{MAX}.
Provided $L/\xi(K)$ is fixed with respect to $K$,
the value itself will in fact not be important 
for the proof of HR,
as can be verified by the theory of finite-size-scaling (FSS), 
i.e., $g_{L}(K)/g(K)=f_{g}(L/\xi(K))$\cite{KIM1}. 
Namely, for a fixed $L/\xi(K)$, $g_{L}(K)$ is exactly
proportional to its bulk value regardless of $K$, 
so that HR can be verified by simply observing 
the constancy of $g_{L}(K)$ with respect to $K$ 
in the  scaling regime.  
This was already illustrated for the 2D Ising model\cite{KIM},
as well as for the 2D XY and Heisenberg models\cite{KIMB}.

BK have not done anything in their Letter `` for the first time,'' 
except probably using the so-called improved estimator of $g(K)$.
With its usage, nevertheless, they did not go deeper into
scaling region than KP, so that it cannot be regarded to be 
of true merit. $g(K_{c})$ was already measured by KP in \cite{KIM}
($g(K_{c})=5.36(7)$ with $K_{c}=0.221650$).

\par
\medskip\noindent
Jae-Kwon Kim \par
Center for Simulational Physics \par
The University of Georgia \par
Athens, GA 30602 \par
}

\end{document}